\documentclass[twocolumn,showpacs,preprintnumbers,amsmath,amssymb,prb,floatfix]{revtex4}

\usepackage{graphicx}
\usepackage{dcolumn}
\usepackage{bm}
\usepackage{natbib}

\begin{document}

\preprint{}

\title{Numerical analysis of the quantum dots on off-normal incidence ion
  sputtered surfaces}

\author{Emmanuel O. Yewande}
\email{e.yewande@mmu.ac.uk}
\thanks{present address: Dept. of Computing \& Mathematics, 
MMU, John Dalton Building, Manchester M1 5GD, United Kingdom.}
\author{Reiner Kree}%
\email{kree@theorie.physik.uni-goettingen.de}
\author{Alexander K. Hartmann}%
\email{hartmann@physik.uni-goe.de}
\affiliation{%
Institut f\"ur Theoretische Physik, Friedrich-Hund Platz 1, D-37077
G\"ottingen, Germany.
}%

\date{\today}

\begin{abstract}
We implement substrate rotation in a 2+1 dimensional solid-on-solid 
model of ion beam sputtering of solid surfaces. With this extension of
the model, we study the effect of concurrent rotation, as the surface
is sputtered, on possible topographic regions of surface patterns. In
particular we perform a detailed numerical analysis of the time
evolution of dots obtained from our Monte Carlo simulations at
off-normal-incidence sputter erosion. We found the same power-law
scaling exponents 
of the dot characteristics for two different sets of ion-material
combinations, without and with substrate rotation. 
\end{abstract}

\maketitle

\section{\label{sec:}INTRODUCTION}
The size-tunable atomic-like properties of (e.g.\ II-VI and III-V)
semiconductor nanocrystals have diverse applications. These properties  
arise from the quantum confinement of electrons or holes in the
quantum dots (QDs) to a region on the order of the electrons' de Broglie
wavelength. Examples of the applications can be found in solid-state quantum
computation \cite{Gershenfeld98} and in 
electronic and opto-electronic devices like   
diode lasers, amplifiers, biological sensors,              
electrolumniscent displays, and photovoltaic cells. \cite{Colvin94, ORegan91}

A more recent method of fabrication is the sputtering of
semiconductor surfaces with a beam of energetic ions impinging at an
angle $\theta$ with respect to the direction perpendicular to the surface. \cite{Facsko99}
This has been 
shown to be a cost-effective and more efficient means of producing
uniform high-density semiconductor nanocrystals,
\cite{Facsko99, Frost00, Facsko01, Gago01} in contrast to previous
methods such as epitaxy, lithographic techniques, colloidal
synthesis, electrochemical techniques, and pyrolytic synthesis.      
  
Using the continuum theory, it was shown that QD formation by $\theta
=$ 0 sputtering is restricted 
to a very narrow region of the parameter space. \cite{Kahng01,
  Frost02} It has also been shown 
that dot formation is possible for $\theta >$ 0, within a broader
region, under concurrent
sputtering and sample rotation.\cite{Frost02} Using a simple discrete
solid-on-solid  
model, which includes the competing processes of surface roughening via
sputtering and surface relaxation via thermal diffusion, we recently
found that for $\theta >$ 0, without sample
rotation, a dot topography is only one among other possible
topographies which may 
arise. The type of the emerging topography 
 depends on the longitudinal and lateral straggle of the
impinging ion as it dissipates its kinetic energy via collision
cascades with atoms within the material. \cite{Yewande06}

In this study, we implement sample rotation in the simulation model
along the lines of Refs.\ \onlinecite{Bradley_Cirlin96} and
\onlinecite{Bradley96}. 
For varying values of the ion parameters we find different kinds of surface
patters, including dots. 
The dots are similar to those obtained without sample rotation, but
the underlying oriented ripple structures are lacking. We study the time
evolution of dots emerging from oblique ion incidence in more 
detail, performing simulations for two different sets of parameters,
corresponding to two different materials (GaAs sputtered with Ar and
Si sputtered with Ne) with and without rotation. We found
that without rotation 
the average number of dots decreases with increasing fluence, but stays
approximately constant for a rotated sample. 
Furthermore the uniformity of dots is
greatly enhanced by sample rotation. Remarkably, both materials exhibit
the same scaling of the dot characteristics with sputter time.

The rest of the paper is organized as follows: In section \ref{sec:CT}
we shall briefly review the continuum theory of sputtered amorphous
surfaces. In section \ref{sec:sim} we shall give a brief description
of the discrete simulation models applied in this 
work. In section \ref{sec:rot_effect} we study the effect of rotation
on the possible topographies reported in Ref.\ \onlinecite{Yewande06};
expecially off-normal incidence dot formation. 
Finally, in section \ref{sec:analysis}, we shall present and discuss our
results of the analysis of dot topographies.       

\section{\label{sec:CT}CONTINUUM THEORY}  
In the seminal theory of P. Sigmund on ion-beam sputtering of amorphous 
and poly-crystalline targets, it was shown that the spatial energy
distribution  
$E({\bf r})$ of an impinging ion may be approximated by a two-dimensional 
Gaussian of widths 
$\sigma$
and $\mu$, parallel ($z'$-axis) and perpendicular 
($x',y'$-axes) to the ion beam direction, respectively
\cite{Sigmund69, Bradley88},   
\begin{equation}
\label{eq:GE} 
E({\bf r}')=\frac{\epsilon}{(2\pi)^{3/2}\sigma\mu^2}\exp\biggl(
-\frac{[z'+a]^2}{2\sigma^2}-\frac{x'^2+y'^2}{2\mu^2}\biggr)\,,    
\end{equation}
where $\epsilon$ is the total energy of an impinging ion and 
$a$ the average penetration depth. 

When describing the surface by a two-dimensional height field $h({\bf r}, t)$
(solid-on-solid description),
the normal erosion
velocity   $v=-[1+(\nabla h)^2]^{-1/2}\partial_t h$  
at the point ${\bf r}=(0,0,0)$ is proportional to the total power
transported to this point by
the ions of the impinging ion beam. It may
be expressed as  
\cite{Cuerno95, Makeev02} 
\begin{equation}
\label{eq:erov} 
v=p_c\int_Rd{\bf r}\Phi({\bf r})E({\bf r}).  
\end{equation}
The integral is taken over the region $R$ containing all the points at
which the deposited energy contributes to the total power at ${\bf r}=0$
and $p_c$ is a proportionality constant.\cite{Makeev02}
 $\Phi({\bf r})$ is a local correction to the uniform flux $J_f$. Note
 that shadowing effects among neighboring points and redeposition of
 eroded material are ignored. 

By standard arguments, an equation for the evolution of the surface height
field $h(x,y,t)$ due to sputtering can be derived from Eq.\ (\ref{eq:erov}), which
takes on the form
\cite{Bradley88, Cuerno95}
 \begin{eqnarray}
\label{eq:hS}
(\partial_th)_S \approx -v_f +
v_f^\prime\partial_xh+\nu_x\partial^2_xh+\nu_y\partial^2_yh 
\nonumber
\\
+\frac{\chi_x}{2}(\partial_xh)^2+\frac{\chi_y}{2}(\partial_yh)^2 +
\eta_S({\bf r}, t),
\end{eqnarray}
where the $x$-axis is parallel to the ion beam direction, $v_f$ is the
erosion velocity of a flat surface and $\nu_x(\theta, \sigma, \mu)$ 
[$\nu_y(\theta, \sigma, \mu)$] is the surface-tension 
coefficient along [perpendicular to] the ion beam
direction. Depending on the parameters $\theta,
\sigma, \mu$,  the quantity $\nu_x$ can exhibit positive as well as negative values, whereas $\nu_y$ is
always negative. $\chi_x$ and $\chi_y$ are the coupling constants of
the dominant non-linearities along the
respective directions. $\eta_S$ is an uncorrelated noise term, with zero
mean. This term represents the random arrival of the ions onto the
surface.     

The surface height also evolves due to surface particles hopping from
one point to the other. On a coarse-grained
level this can be described  by the continuity equation  
for the conserved particle current  
\begin{equation}
\label{eq:hD}
(\partial_t h)_D=-\nabla\cdot {\bf j}+\eta_D({\bf
  r}, t),
\end{equation}  
where the local current density  ${\bf j}$ is a function of the
derivatives of $h$; i.e, ${\bf j}$ = ${\bf j}[\nabla^m h, (\nabla
h)^n]$ ($m$ and $n$ are integers), and the acceptable
functional form is subject to symmetry constraints. $\eta_D$ reflects the
randomness inherent in the surface-diffusion process.                  
A commonly studied example of a surface migration model is the
Mullins-Herring model, 
which leads to the particularly simple form
 $\nabla\cdot {\bf j}=D\nabla^4h$ for the current density.\cite{Kim94}

Thus, considering Eqs.\ (\ref{eq:hS}) and (\ref{eq:hD}), the time evolution
of a sputtered surface is governed by a Kuramoto-Sivashinsky type
equation 
\begin{eqnarray}
\label{eq:KS}
 \partial_th = -v_f +
v_f^\prime\partial_xh+\nu_x\partial^2_xh+\nu_y\partial^2_yh 
+\frac{\chi_x}{2}(\partial_xh)^2 
\nonumber
\\
+\frac{\chi_y}{2}(\partial_yh)^2
-D\nabla^4h +
\eta({\bf r}, t).
\end{eqnarray} 

The exact form of each coefficient of
Eq.\ \ref{eq:KS} is given in Ref.\ \onlinecite{Makeev02}
and will be used later in the analysis of our
results (see Fig.\ (\ref{fig:iso_coeff}) in section\ (\ref{sec:sim})).

The anisotropy ($\nu_x\ne\nu_y$, $\chi_x\ne\chi_y$) arises from the
oblique incidence which leads to different erosion rates (or different
rates of maximizing the exposed area) along
parallel and perpendicular directions relative to the ion beam
direction. 
From the linearized Eq.\ (\ref{eq:KS}) it is easy to see that periodic ripples are growing  with wavevector ${\bf
k}=\sqrt{\max(|\nu_{x/y})|/2D}$ driven by the dominant negative
surface tension. These ripples are stabilized by the surface diffusion term.
With prolonged
sputtering surface slopes become too big to neglect the nonlinear terms and the ripples are
destroyed, after which a new rotated ripple structure may
emerge. \cite{Park99} 

If the sputtered substrate is rotated with sufficiently large angular
velocity, \cite{Thesis} 
or if $\theta=$ 0, no externally induced 
relevant anisotropy is present. 
In this case
ripples do not develop for the class of materials we study here 
and the relevant evolution equation is the
isotropic version of Eq.\ (\ref{eq:KS}), which predicts cellular
structures of mean width $\sim (D/|\nu|)^{1/2}$ and mean height $\sim
|\nu|/|\chi|$.\cite{Bradley96} According to Ref.\
\onlinecite{Kahng01}, 
these cellular structures eventually evolve to either a dot 
topography ($\chi>$0) or a hole topography ($\chi<$0); which explains
the normal incidence dot\cite{Facsko99, Gago01, Facsko01} and
hole\cite{Rusponi98, Rusponi99} topographies found in 
previous experiments.    

Thus, one would expect oblique incidence dot formation to result from 
isotropic sputtering induced by substrate rotation. However,
in our recent simulations \cite{Yewande06} 
we found dot topographies without sample rotation,
though with some 
underlying (anisotropic) ripples oriented parallel to the
ion beam direction. We found that the oblique incidence dots arise
from a dominance of the ripples oriented parallel to the ion beam
direction, over those with a perpendicular orientation; the dots being the
remains of such perpendicular ripples (see Fig.\
\ref{fig:phase_profiles} below) after long times.  

In the next chapter
we will outline the simulation method we have used in 
Ref.\ \onlinecite{Yewande06} and which we have extended to the case
with rotation as presented in this work.

\section{\label{sec:sim}MONTE CARLO SIMULATION}

\begin{figure}[htb]
 \begin{center}
 \includegraphics[width=0.8\columnwidth]{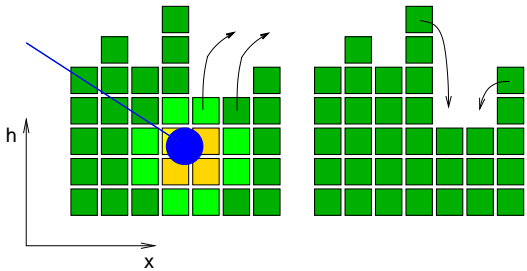}
 \end{center}
 \caption{
 \label{fig:Model}
The model consists of a square field  of discrete height
  variables $h(x,y)$, corresponding to piles of $h(x,y)$ particles at position
  $(x,y)$. Here a projection to the $h-x$ is shown. 
{\em Left:} Each ion impact is modeled by an distribution describing
  the energy deposited by the ion. Atoms on the surface are removed
  with a probability proportional to the energy. {\em Right:} Surface
  diffusion; by decreasing height differences the energy is decreased.
}
 \end{figure}

To simulate the competing processes discussed in section \ref{sec:CT}
on a discrete system, we use the 
Monte Carlo model of sputter erosion introduced in Ref.\
\onlinecite{Hartmann02} (HKGK model). To make the paper self-contained,
we provide some details here. 
We simulate the sputtering process on an initially flat surface of size $L^2$ with periodic
boundary conditions, by starting an ion at
a random position in a plane parallel to  the initial 
surface, and projecting it along a straight trajectory inclined at
angle $\theta$ to the direction perpendicular to the averaged surface
configuration and at an azimuthal angle
$\phi$. An ion penetrates the solid through a
depth $a$ and releases its energy according to Eq.\ (\ref{eq:GE}). 
An atom at a position $(x,
y, h)$ is eroded (see Fig.\ \ref{fig:Model}) with probability
proportional to $E({\bf r'})$ 
given in Eq.\ (\ref{eq:GE}). It should be noted that,
in accordance with the assumptions of the theoretical models 
\cite{Makeev02, Bradley88, Cuerno95}, this
sputtering model neglects evaporation, redeposition of eroded
material and preferential sputtering of surface material at the point of
penetration. The surface is defined by a single valued, discrete time
dependent SOS height function $h({\bf r},t)=h(x, y, t)$.  
The time t is measured in terms of the
ion fluence; i.e, the number of incident ions per two-dimensional 
lattice site $(x,y)$. The choice of the parameters
$\sigma,\mu$ and $\theta$ is discussed below.
Substrate rotation, at constant angular
velocity $\omega\ge \nu^2/D$, is implemented by keeping the substrate static and
then rotating  
the ion beam instead, which is equivalent to keeping the ion beam direction fixed 
and rotating the substrate \cite{Bradley_Cirlin96}. In our
simulations we choose
$\phi$ randomly 
from the interval $0 \le \phi < 2\pi$, since, in the large $\omega$ limit the solid appears  
to be sputtered from all angles $\phi$.\cite{Bradley96}

Our model of the sputtering mechanism sets the time scale of the
simulation in a way, which allows direct comparison with experiments. 
Any relevant surface diffusion
mechanism may be combined with this sputtering model. \cite{Thesis}
Here, we use a realistic solid-on-solid model of thermally activated surface
diffusion \cite{Smilauer93}. 
Surface diffusion is simulated as a nearest neighbor
hopping process with an Arrhenius hopping rate
\begin{equation} 
\label{eq:arrhenius}
R(E, T)=R_0\exp(-E/k_BT),
\end{equation}
where $R_0=k_BT/\hbar$, and $T$ is the {\em effective} substrate
temperature (see below).
 The 
energy barrier $E=E_{vn}+n_nE_{ln}+E_{se}$ consists of a
substrate term ($E_{vn}=$0.75 eV), a nearest neighbor term
($E_{ln}=$0.18 eV), and a step barrier term ($E_{se}=$0.15 eV), which
only contributes in the vicinity of a step edge, and is zero
otherwise.  
In each diffusion sweep all
surface particles are considered, and those that are not fully
coordinated may hop to a neighboring site. Note
that we have to use a higher effective temperature \cite{Yewande05}
in our simulation
in order to account for the greatly enhanced surface diffusion due to {\em thermal spikes}. 
A thermal spike is a series of sharp peaks and
minima in the spatio-temporal distribution of the surface temperature,
arising from the occurence of local heating of the surface right after every
ion impact, followed by rapid cooling. Hence, we have used a higher
effective temperature $k_BT=$0.1 eV, which was estimated in our
previous work and which roughly corresponds to experimental sputtering at room
temperature. 

The model captures the essential features of sputtered surfaces at
nanometer lengthscales; especially nonlinear effects
\cite{Hartmann02, Yewande05}.    
Although a direct mapping of this model to the continuum equations is
unavailable, the model is expected to be consistent with variants of the KS
equation. \cite{Yewande06, Lauritsen96} 

\section{\label{sec:rot_effect}EFFECT OF ROTATION ON THE 
TOPOGRAPHIES}

In this study, a lattice of linear size  $L$ =
128 is used with a lattice spacing corresponding roughly to a distance
of 0.5 nm. The simulated time is set such that
1 ion/atom corresponds to an ion fluence of
3$\times$10$^{14}$ ions/cm$^2$.  
We use a sputter yield of about 7 surface atoms/ion, as
compared to 5 SiO$_2$ molecules/ion (reduced to 0.1 SiO$_2$
molecules/ion with H ion) and 0.3-0.5 molecules/ion in the experiments
of Refs.\ \onlinecite{Mayer94} and \onlinecite{Habenicht99}
respectively; this may result in lengthscales differing from
those of the cited experiments, but we found in our previous studies 
\cite{Hartmann02,Yewande05,Yewande06} that predicted  
universal features exist, which are directly comparable to
experimental results. We will subsequently discuss such features for
rotated samples.  

\begin{figure}[!htbp]
\begin{center}
\includegraphics[width=0.48\textwidth]{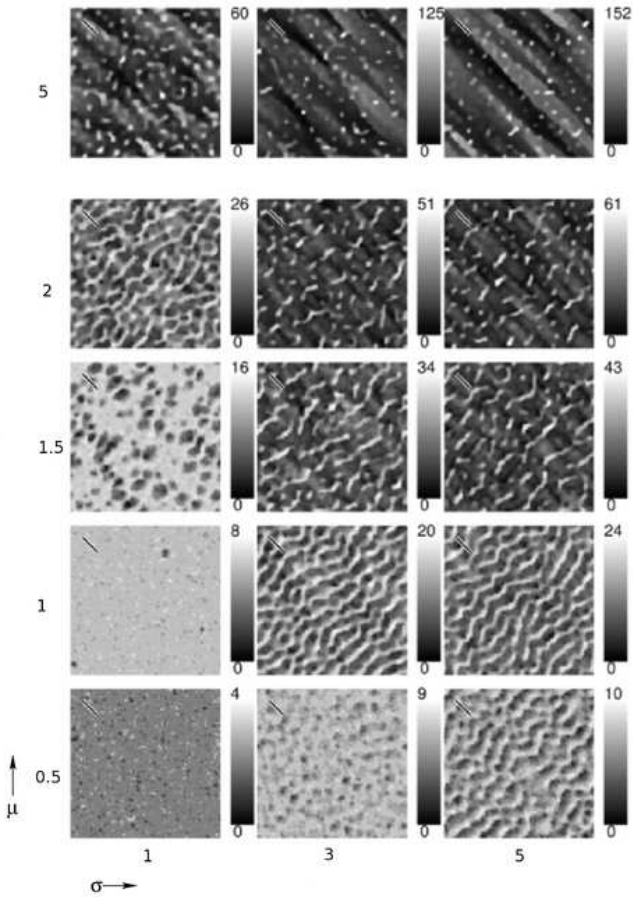}
\caption[Phase profiles,
t=3]{\label{fig:phase_profiles}  Possible topographies of the model,
  within the experimental constraints considered in Ref.\
  \onlinecite{Yewande06}. $t =$ 3, $a =$ 6, $\theta =$
  50$^\circ$. Left - right columns: $\sigma$ = 1, 3, and 5,
  respectively.   
Bottom row - top row: $\mu$ = 0.5, 1, 1.5, 2, and 5, respectively. 
The last two profiles of the top row belong to the dot
region (region V) of Ref.\ \onlinecite{Yewande06}.}
\end{center}
\end{figure}

\begin{figure}[!htbp]
\begin{center}
\includegraphics[width=0.48\textwidth]{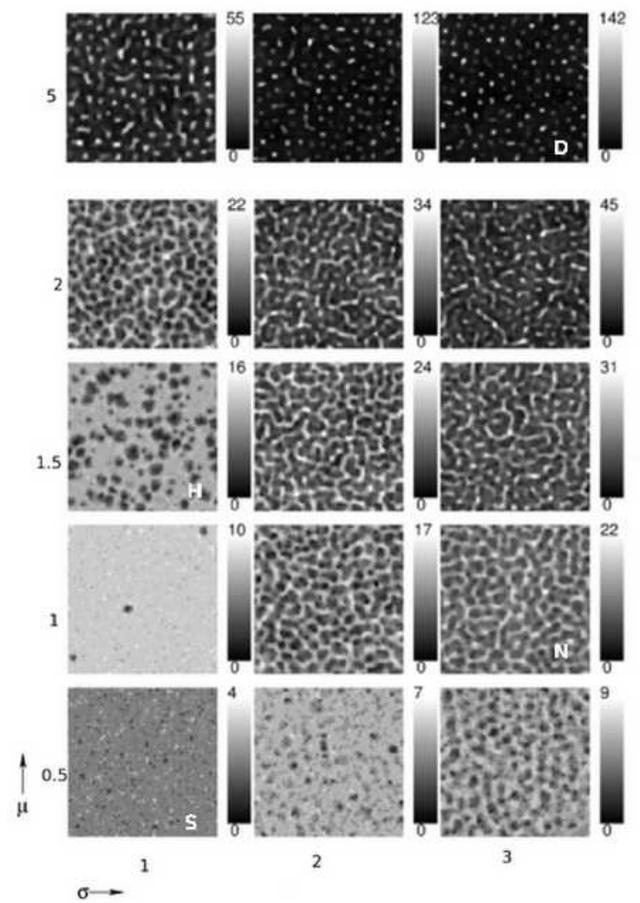}  
\caption[Concurrent sputtering and rotation (PSSR),
t=3]{\label{fig:phase_profiles_rot}  Profiles obtained from simultaneous 
  sputtering and rotation, using the same parameters as in Fig.\
  \ref{fig:phase_profiles}. $t =$ 3, $a =$ 6, $\theta =$ 50$^\circ$.  
Left - right columns: $\sigma$ = 1, 3, and 5,
respectively.   
Bottom row - top row: $\mu$ = 0.5, 1, 1.5, 2, and 5, respectively. 
The last two profiles of the top row belong to the dot
region (region V) of Ref.\ \onlinecite{Yewande06}. Structure factors
of the lettered profiles are provided in Fig.\ \ref{fig:Sk_rot} [S -
(relatively) smooth; H - hole; N - non-oriented structures; D - dot].} 
\end{center}
\end{figure}

In Ref.\ \onlinecite{Yewande06} six regions with different topographies
(including e.g, smooth, hole, and dot regions) were shown to emerge for
ion collisional parameters $a$ = 6, 0 $\le \sigma \le$ 5, 0 $\le \mu
\le$ 5 (all measured in units of lattice spacings), 
at time $t$ = 3 ions/atom. In this section we study
the effects of sample rotation 
on these topographies, in particular  effects on the region with dot topography.
The differences in topographies between rotated and unrotated samples
are visualized in Fig.\
\ref{fig:phase_profiles} and Fig.\
\ref{fig:phase_profiles_rot}, for $a$ = 6, $t$ = 3, and $\theta$ =
50$^\circ$. 

As seen from Fig.\ \ref{fig:phase_profiles_rot} (and Fig.\
\ref{fig:Sk_rot}, see below), no anisotropy can be found with substrate 
rotation, as expected from the continuum theory. The ripple structures obtained
for $\mu \le$ 2 (Fig.\ \ref{fig:phase_profiles}) do not appear for rotated
substrates. The underlying parallel ripples of the dot
region (topmost row of Fig.\ \ref{fig:phase_profiles}) are also
absent for rotated substrates. However, hole formation is not 
suppressed,  
we get holes  with as without rotation as is visible  in Fig.\
\ref{fig:phase_profiles}.
This fact can be understood from the continuum theory,
which predicts roughly equal 
erosion rates along both directions for parameters in the hole
region,\cite{Yewande06} hence there is no anisotropy to be 
destroyed. Furthermore, 
ripple patterns perpendicular with resepct to the ion beam direction are replaced by  
non-oriented structures, and the ordered parallel ripples are 
no longer present if the substrate is rotated.  
(see Fig.\ \ref{fig:phase_profiles_rot}), 

For a closer inspection, we calculate the 
structure factors, $S({\bf k}, t)=|h({\bf k}, t)|^2$  from the Fourier
transform $h({\bf k}, t)$ of the height field $h({\bf r},t)$. In
particular we consider four prototypical topographies marked by
letters  S, H, N, D in Fig.\ 
\ref{fig:phase_profiles_rot}.  S stands for ``relatively smooth'', H for ``hole'', N
for ``non-oriented structures'', and D for ``dots''. The results are shown in Fig.\
\ref{fig:Sk_rot}. As can
be seen from this figure, and as expected, there is no anisotropy
visible in
all cases. In the 
case of the relatively smooth surface S, there is also no 
characteristic lengthscale. For the hole
topography, H, there is still no specific lengthscale but there now
exists an upper bound $k_{ub}$ on $|{\bf k}|$ due to the presence of
the holes. On the surface with non-oriented
structures (N) a well defined lengthscale with $k_{ub}$ as well as a lower
bound $k_{lb}$ can be found. And finally, in the case of the dot topography (D), we
also have a characteristic lengthscale, but $k_{lb}$ is shorter here
than for the N topography, which implies
that the average separation of the dots is larger than that of the
non-oriented structures, as expected from Fig.\
\ref{fig:phase_profiles_rot}.

\begin{figure}[!htbp]
\begin{center}
\includegraphics[width=0.5\textwidth]{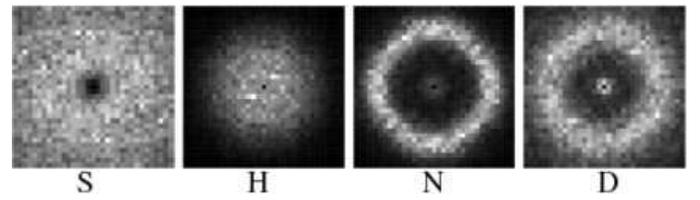} 
\caption[Structure factor]{\label{fig:Sk_rot}  Structure factor
  of the lettered surface profiles in Fig.\
  \ref{fig:phase_profiles_rot}. S $\Rightarrow$ 
(relatively) smooth; H $\Rightarrow$ hole; N $\Rightarrow$
non-oriented structures; D $\Rightarrow$ dot. ${\bf k}={\bf 0}$ at 
the centre; and ${\bf k}=\frac{2\pi}{8}(\pm 1, \pm 1)$ at the corners.}
\end{center}
\end{figure}

\begin{figure}[!htbp]
\begin{center}
\includegraphics[width=0.99\columnwidth]{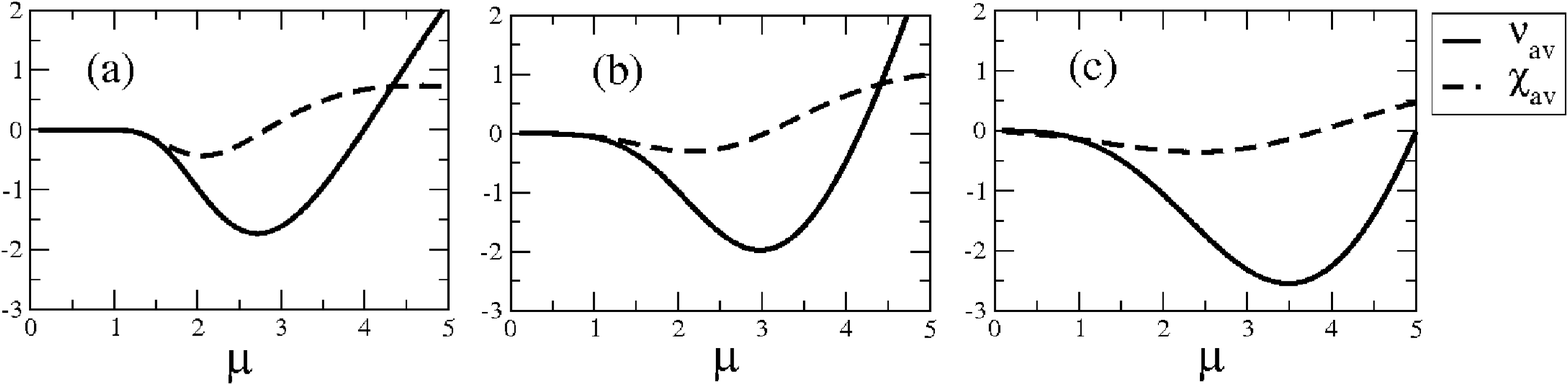} 
\caption[Coefficients of the iKS equation for the rotated
case]{\label{fig:iso_coeff}  The coefficients, $\nu_{av}$ and 
  $\chi_{av}$, of the isotropic version of Eq. \ref{eq:KS}, 
  \cite{Bradley96} for the  
rotated case, as functions of $\mu$. (a) $\sigma =$
 1.0, (b) $\sigma =$ 3.0, and (c) $\sigma =$ 5.0.} 
\end{center}
\end{figure}
 
According to the continuum theory, there exists a single effective surface tension 
coefficient $\nu_{av}$ = $\nu_x$ + $\nu_y$ and a single nonlinear
coupling\cite{Bradley96} $\chi_{av}$  for all directions  in rotated samples,
since there is no anisotropy left in the system. 
Fig.\ \ref{fig:iso_coeff} shows a plot of these coefficients based
upon the explicit expressions in Ref.\ \onlinecite{Makeev02}. 
As can be seen from Fig.\ \ref{fig:iso_coeff} and Eq.\ (\ref{eq:KS}) 
the surface roughens with
time, with smaller  
$\nu_{av}$ ($\nu_{av}<$0) corresponding to higher roughness. 
Surfaces in the parameter range for which 0$> \nu_{av} \gg$ -1, are
relatively smooth (S) topographies.

\begin{figure}[!htbp]
\begin{center}
\includegraphics[width=0.4\textwidth]{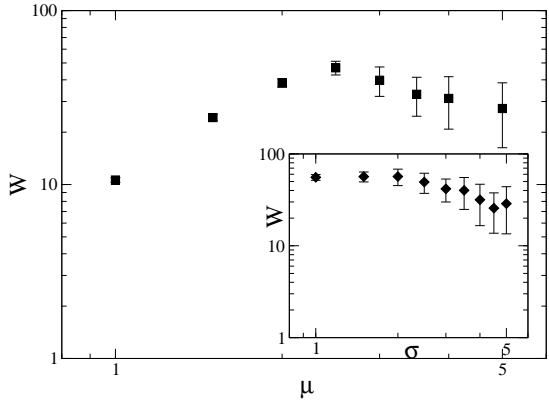} 
\caption[Roughness as a function of collision cascade
parameters]{\label{fig:roughness}  Surface roughness, $W$, with dots
  excluded. Main plot: $W$ as a function of $\mu$ for $\sigma=$ 5. 
  Inset: $W$ as a function of $\sigma$ for $\mu=$ 5. $t=$ 3 ions/atom.   
} 
\end{center}
\end{figure}

Considering Fig.\ \ref{fig:iso_coeff}, one sees that  
$|\nu_{av}|$ first increases, in accordance with
the increasing height difference on the  
greyscale charts on the profiles of Fig.\
\ref{fig:phase_profiles_rot} and then it decreases as we
tend towards the dot region. These changes 
in the roughness are not visible on the greyscale charts due to the
appearance of the dots, which are considerably higher
than an average surface protrusion in the dot-free profiles. 
Therefore we have also studied the surface roughness, with the dots
excluded, as shown in Fig.\ \ref{fig:roughness} (Details of our 
dot-isolation method are discussed in the next section).  In
this figure, the 
roughness $W$ as a function of $\mu$ ($\sigma=$ 5) is 
shown in the main 
plot, where the roughness first increases and than decreases again,
in accordance with the continuum theory. The inset shows a plot of $W$ versus
$\sigma$, for $\mu=$ 5. 
 
\begin{figure}[!htbp]
\begin{center}
\includegraphics[width=0.3\textwidth]{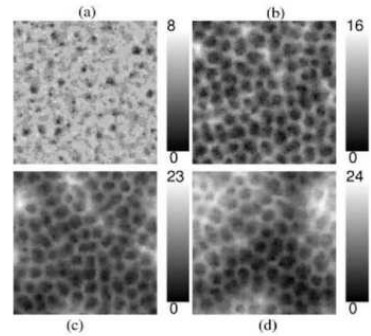} 
\caption[Time evolution of rough surface with sample
rotation]{\label{fig:rough_tevol_rot}  Time evolution of the
  relatively smooth topography of Fig.\ \ref{fig:phase_profiles} 
  with rotation. $\sigma =$ 3, $\mu =$ 0.5. 
(a)-(d): $t =$ 3, 40, 90, and 150, respectively.} 
\end{center}
\end{figure}

 When the local surface slopes become significant (with prolonged
sputtering), nonlinearities become relevant. It has been shown that for $\theta =$ 0 crossover to the nonlinear regime either gives
rise to dot formation (if $\chi_{av} >$ 0) or hole formation (in case
$\chi_{av} <$ 0)   if 
ion-induced effective surface diffusion is the dominant relaxation
mechanism  \cite{Kahng01}. This is consistent with
our results for parameters at which
$\chi_{av} >$ 0 in Fig.\ \ref{fig:iso_coeff} (i.e, for $\mu \gtrsim$
3). We also found holes for $\chi_{av} <$ 0 (``H'' region), but hole formation for long times is not as widespread as
Fig.\ \ref{fig:iso_coeff} seems to indicate.
In particular, the hole topography eventually evolves into cellular
structures similar to those shown in
Fig.\ \ref{fig:rough_tevol_rot}  at long times. 
Since $\nu_{av} \ne$ 0, the surface roughening is not 
wavelength independent, which explains the presence of the non-oriented protrusions. 

For very small longitudinal and lateral straggle, $\sigma\le$
1, $\mu\le$ 0.5, i.e. in the ``S'' region, 
we did not find any structure up to the longest simulation times. 
Note that with increasing $\sigma$, the $|\nu_{av}| \approx$ 0
interval is reduced.

\begin{figure}[!htbp]
\begin{center}
\includegraphics[width=0.3\textwidth]{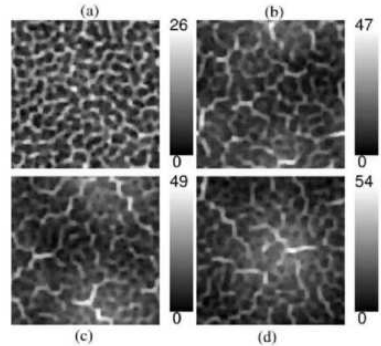} 
\caption[Time evolution a non-oriented structure with sample
rotation]{\label{fig:ripple_tevol_rot}  Time evolution of the 
  non-oriented structures  arising from rotation of rippled region 
  of Fig.\ \ref{fig:phase_profiles}. $\sigma$ = 3, $\mu$ = 1.5. (a) -
  (d): $t$ = 3, 40, 
  90, and 150, respectively.} 
\end{center}
\end{figure}

For non-oriented structures ``N'', simulations at longer times reveal
only slight  changes in the structures; no dot formation (see 
Fig.\ \ref{fig:ripple_tevol_rot}) appears.
Hence, we have only observed dots with rotation wherever they can be
found without rotation.

 \begin{figure}[!htbp]
\begin{center}
\includegraphics[width=0.5\textwidth]{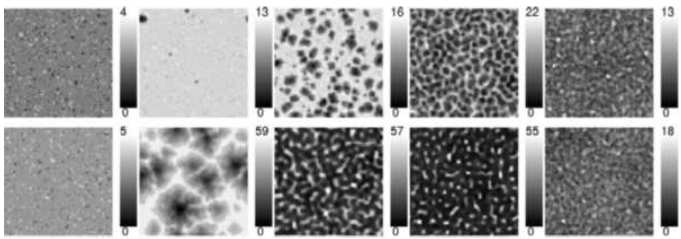} 
\caption[Topographic transitions for $\sigma=$ 1, $\mu=$ 2 and
5]{\label{fig:diff_theta}  Topography changes as a result of
  changing the angle of incidence $\theta$, for $\sigma=$ 1; $\mu=$ 2
  (top row), and 5 (bottom row). Top row, L-R: $\theta=$ 10, 30, 40,
  50, 80. Bottom row, L-R: $\theta=$ 5, 10, 30, 50, 80.} 
\end{center}
\end{figure}

\begin{figure}[!htbp]
\begin{center}
\includegraphics[width=0.4\textwidth]{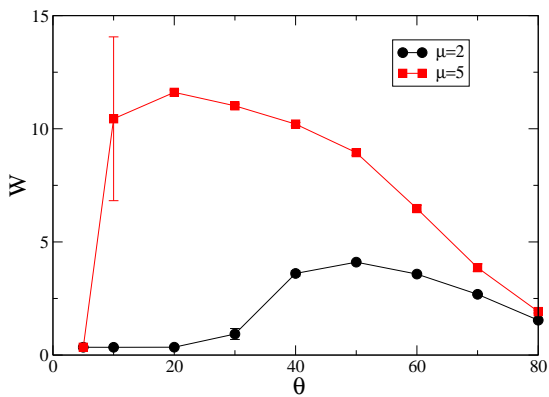} 
\caption[Roughness as a function of collision cascade
parameters]{\label{fig:rough_diff_theta}  (Color online) Surface
  roughness      
  $W$ as a function of $\theta$ for profiles shown in Fig.\
  \ref{fig:diff_theta}. Data for $\mu$ = 2 and 5 are represented by
  circle and square symbols, respectively. In both cases $\sigma$ = 1,
  $t$ = 3 ions/atom. Error bars are included, but they are much smaller than
  symbol size except at $\theta=10^{\circ}$ ($\mu=5$). For this case
  for about half of the runs the structure had already evolved to a
  non-oriented structure pattern with a considerable roughness, 
while the other half exhibited still  a rather smooth surface. Hence,
there seems to be a sharp transition in time from smooth to rough,
where the transition time of finite samples fluctuates strongly.
} 
\end{center}
\end{figure}

When considering the emerging topographies of Fig.\
\ref{fig:phase_profiles_rot} at other angles of incidence, we found
notable changes with $\theta$, as illustrated in Fig.\ \ref{fig:diff_theta} for $\sigma=$ 1, $\mu=$ 2 and $\sigma=1$, $\mu=5$). 
That is, changes from
smooth $\to$ 
hole $\to$ non-oriented structure, and back to smooth (no structure)
topographies appear. These changes imply that the roughness 
initially increases and then starts to decrease with 
increasing $\theta$ as shown in
Fig.\ \ref{fig:rough_diff_theta}. In this figure $\sigma$ = 1 and data
for $\mu$ = 2 and $\mu=5$ are represented by  
  (black) circle and (red) square symbols, respectively. The 
  behavior of the roughness with 
varying $\theta$ shown in Fig.\ \ref{fig:rough_diff_theta} is in
agreement with the experiments reported for Ar-ion 
sputtered rotating InP surfaces in Ref.\
\onlinecite{Frost00}. But note that the roughness data reported in this
experiment \cite{Frost00} were obtained for an exposure time of 1200
secs., i.e.\, in the steady state regime. To use such steady state
values in Fig.\ \ref{fig:rough_diff_theta} would require simulation
times beyond our computational time constraints.

\section{\label{sec:analysis}NUMERICAL ANALYSIS OF THE DOTS}

\begin{table}
\caption{\label{tab:SRIM}SRIM results for materials bombarded with
  noble gas ions.}
\begin{ruledtabular}
\begin{tabular}{llllll}
Ion & Material & E/keV & $a$ \footnotemark[1]\footnotetext[1]{in
  lattice units} & $\sigma$ \footnotemark[1] & $\mu$ \footnotemark[1] \\
Ar & GaAs & 1.6 & 6 & 5 & 3.8 \\
Ar & Ge & 1.6 & 6 & 5 & 3.8 \\
Ne & Si & 0.65 & 6 & 4.4 & 3.2 \\
\end{tabular}
\end{ruledtabular}
\end{table}

Focussing on topographic region ``D'' where dots have
been found also without 
rotation, we obtained the experimental parameters corresponding to
this region from SRIM simulations \cite{srim} as shown in Table
\ref{tab:SRIM}. We have performed two sets of simulations using the
SRIM results: (i) 1.6 keV Ar ion sputtering of GaAs surfaces; where  
$a$ = 6, $\sigma$ = 5, $\mu$ = 3.8. (ii) 650
eV Ne ion sputtering of Silicon surfaces; where $a$ = 6, $\sigma$ =
4.4, $\mu$ = 3.2. In both cases we have used 
an ion beam inclination of $\theta=$50$^\circ$ to the vertical. 

In order to study the time evolution of dot characteristics
(e.g. dot density, area, and height), we have used a similar clustering
approach as in Ref.\ \onlinecite{Yewande05}. A dot is defined as a
cluster of points of local height maxima.  This is done in two steps
\begin{itemize}
\item For any given time
$t$ (we do not mention the time-dependence explicitly here)
we consider the set of points 
\begin{equation}
M\equiv \{(x, y)|x, y \in (1, \cdots, L); h(x,y)\ge h_c \},
\end{equation}    
where $L$ is the linear size of the system,  $h_c$
is a cutoff height which the surface height at a point must equal/exceed for
the point to be counted as (or part of) a dot. \cite{Yewande05} 
We define $h_c$ to be a function of the average surface protrusion, which
has the form: $h_c=h_{min}+p(\langle h\rangle-h_{min})$. Where
$h_{min}$ is the lowest surface height, $\langle h\rangle$ is the average
surface height, and $p$ is a fixed percentage. 
\item  We call two
  points in $M$  {\em neighbors} if their distance is smaller than a given
  threshold $d_c$, we use $d_c=1$ here. Then, two points are
  called {\em connected}, if there exists a path from the 
first point to the other
  point such that
all consecutive points along the path are neighbors.
Now, each dot $D$ is a subset of $M$ (non-overlapping with
  any other dot) of maximum size such that all points in $D$ are mutually
  connected. Hence, dots are
the transitive closures of the neighbor relation on $M$.
\end{itemize}

\begin{figure}[!htbp]
\begin{center}
\includegraphics[width=0.4\textwidth]
{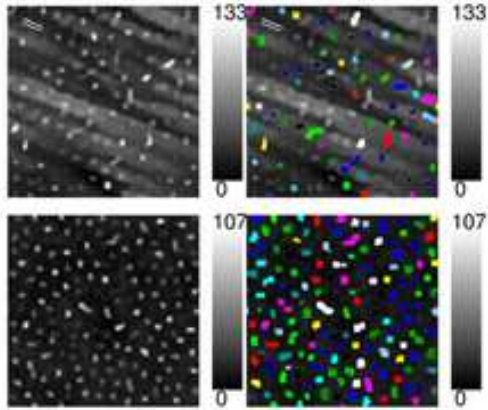}
\caption[clusters for the non-rotated and the rotated
case]{\label{fig:clusters}  (Color online) 
Sample surface profiles for the
  unrotated case (top), and the rotated case (bottom) at t =
  3 ions/atom. In the figures on
  the right the clusters formed from the corresponding profile on the
  left (as defined in the text) are printed on top of the profile. 
  The scales indicate the surface height measured 
  from the lowest height. 
}
\end{center}
\end{figure}

\begin{figure}[!htbp]
\begin{center}
\includegraphics[width=0.5\textwidth]
{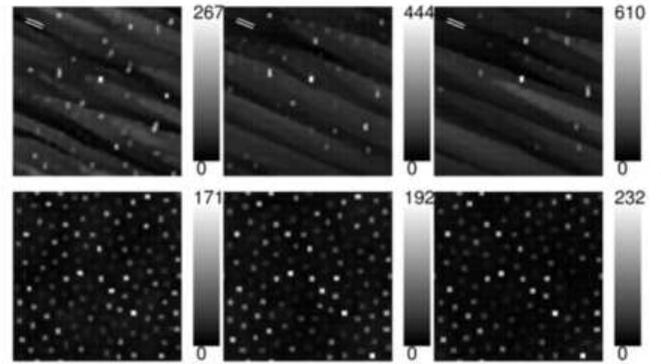} 
\caption[time evolution of the dots the non-rotated and the rotated
case]{\label{fig:dotwt}  Surface profiles for the
  non-rotated (top row) and the rotated (bottom row) case. L - R, t =
  10, 20, and 30 ions/atom respectively. The bar indicates the ion
  beam direction, and the scales indicate the surface height measured
  from the lowest height.  
}
\end{center}
\end{figure}

We start our
simulations with a dot configuration obtained from a topography at $t=$ 3 ions/atom and we choose a
value $p=p_o$ that yields the highest number of sampled dots $N_o$ (see
Table \ref{tab:clustering}).         
The initial clusters are shown in Fig.\ \ref{fig:clusters} for simulation
 of Ar-ion sputtering of GaAs 
without substrate rotation  and with substrate rotation, respectively. 
 The dot cross-section $A(D)$ is defined $A(D)=|D|$; i.e the cardinality of
$D$. The average height $h_d$ of a single dot $D$ is defined as  
$h_d (D) = \frac{1}{|D|}\sum_{(x,y)\in D} h_{x, y}$.  
Note that, in order to exclude non-dot surface  protrusions (see top row of
Fig.\ \ref{fig:dotwt}) from our analysis, we have included an upper
boundary of 50 cluster points (i.e $A(D)\le$ 50) to our definition.
In our 
analysis we only consider dots defined by these clusters.
 The results reported
here are obtained from an average of 100 independent runs.       

  \begin{table}
\caption{\label{tab:clustering} Optimal parameters used to determine
  the cutoff height $h_c$}
\begin{ruledtabular}
\begin{tabular}{llll}
Simulation & $p_o$ (\%) & $N_o$ & State \\
Ar$^+$ on GaAs & 50 & 69 & unrotated \\
Ar$^+$ on GaAs & 10 & 149 & rotated \\
Ne$^+$ on Si & 60 & 76 & unrotated \\
Ne$^+$ on Si & 20 & 139 & rotated \\
\end{tabular}
\end{ruledtabular}
\end{table}

In Fig.\ \ref{fig:dotwt}, we show sample surface profiles without
rotation (top row), and with rotation (bottom row), for simulation of
1.6 keV Ar-ion sputtering of GaAs. As can be seen from the top row of
this figure, the ripples (with parallel orientation to ion beam
direction) that coexist with the dots become more ordered with time,
whereas the dots decrease in number with time; more analysis of these
dots is provided below. On the other hand (bottom row of Fig.\
\ref{fig:dotwt}), these ripples do not exist when the substrate is
subjected to concurrent rotation, and the density of the dots is more
uniform. 

\begin{figure}[!htbp]
\begin{center}
\includegraphics[width=0.4\textwidth]
{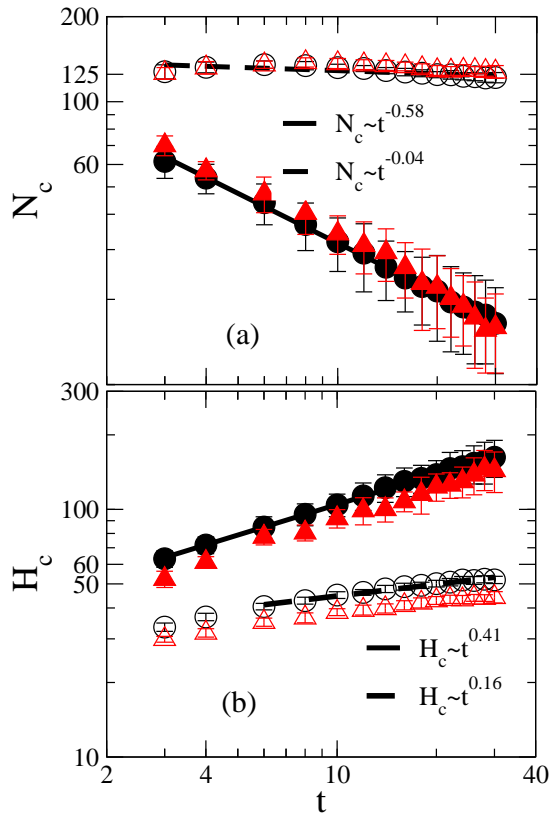}
\caption[time evolution of the dot
characteristics]{\label{fig:dotcharNH}  (Color online)  
Time evolution of (a) the average number N$_c$, and, (b) the average
height H$_c$ (both in lattice units), of 
the sampled dots with time (in ions/atom). Symbols: open and closed
circle (triangle) symbols denote data obtained 
with and without substrate rotation respectively, for Ar-GaAs (Ne-Si).
}
\end{center}
\end{figure}

\begin{figure}[!htbp]
\begin{center}
\includegraphics[width=0.4\textwidth]
{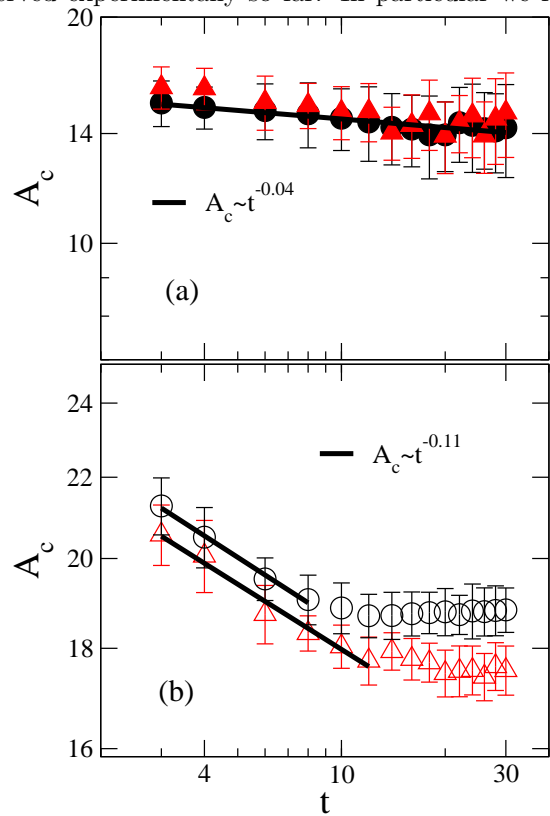}
\caption[time evolution of the dot area]{\label{fig:dotcharA} 
  (Color online)  
Time evolution of the average area of crosssection A$_c$ of the sampled dots
with time for (a) the unrotated case, and (b) the rotated
case. Symbols: open and closed circle (triangle) symbols denote data obtained 
with and without substrate rotation respectively, for Ar-GaAs (Ne-Si).
}
\end{center}
\end{figure}

 In Fig.\ \ref{fig:dotcharNH}, we present results of the average
number of sampled dots N$_c$, and the average dot height
H$_c=\overline{h_d}$, where the average is taken over all dots and all
independent runs. The
results of the average area of cross-section of the dots is presented
in Fig.\ \ref{fig:dotcharA}. 
 In Figs.\
\ref{fig:dotcharNH} and \ref{fig:dotcharA}, open and closed circle
symbols represent data obtained from Ar-GaAs with and without rotation
respectively; while open and closed triangle symbols represent data
obtained from Ne-Si with and without rotation 
respectively.  
The main result is that we found
power law scaling of the dot characteristics with time; with or
without rotation. In addition, we found the same scaling behavior for
the two sets of simulation performed; i.e, 1.6 keV Ar ion sputtering
of GaAs (Ar-GaAs), and 650 eV Ne ion sputtering of Si (Ne-Si).

To be more specific, 
as can be seen from Fig.\ \ref{fig:dotcharNH} (a), the average number
of sampled dots 
decreases with time as N$_c\sim t^{-\psi}$ for non-rotated samples, 
where $\psi=$ 0.583$\pm$0.007, while it stays approximately constant
for rotated samples ($\psi\approx 0$). 
This indicates that the number of dots becomes insignificant without substrate
rotation, so that one should use
rotation if dot creation is the main purpose of the sputtering.
Note that this result is already visible in 
Fig.\ \ref{fig:dotwt} in qualitative form.
   
On the other hand, as shown in
Fig.\ \ref{fig:dotcharNH} (b), the average height
H$_c$ increases with time as H$_c\sim t^\varrho$, the increase is more
rapid for the unrotated case ($\varrho=$0.409$\pm$0.004) than for the
rotated case ($\varrho=$0.159$\pm$0.007). As expected, the dot height
is lower with sample rotation due to the enhanced smoothening effect
of the rotation. \cite{Bradley_Cirlin96, Bradley96} This smoothening
also explains the higher initial number of dots for the rotated case
[Fig.\ \ref{fig:dotcharNH} (a)]: 
The dots that are
not ``visible'' in the unrotated case, since they do not surmount the ripples,
become ``visible'' in the rotated case since the rotation
prevents ripple formation.   

The mean cross sectional area A$_c$ finally becomes almost time independent for
both cases. However, as shown in Fig. \ref{fig:dotcharA} , an initial
power law scaling A$_c\sim t^{-\varsigma}$ occurs, which is more
significant for the rotated case (where $\varsigma$ = 0.113$\pm$0.006)  
than for the unrotated case (where $\varsigma\approx$ 0). 

These results indicate that while no new dots are created with
rotation (for $\theta >$ 0$^\circ$), the uniformity of the density of
the existing dots and the stability of the dot height are greatly
enhanced with substrate rotation. Our results of the analysis of the
dots are in agreement with previous experiments on dots
obtained from normal incidence sputtering of semiconductors (Si),
\cite{Gago01} 
semiconductor alloys (GaSb, InSb); \cite{Facsko99, Facsko01} as well
as oblique incidence dots 
obtained from simultaneous sputtering and sample rotation
(InP). \cite{Frost00} In these experiments, the dot height has been
reported to increase with time; and the average dot size have
been reported to become constant with time.  

To summarize, we have implemented substrate rotation for a
solid-on-solid model of surface sputtering and used it to study the
effects of concurrent rotation on the different possible
topographies. In particular, we have studied the effect of rotation on
the dot region as well as a detailed analysis of the time evolution of
the dot characteristics (number, cross sectional area, and height)
with and without substrate rotation. We found that different materials
whose sputtering parameters fall within this region exhibit the same
scaling behavior. The number of dots $N_c$ formed in the absence of
substrate rotation decrease with time as $N_c\sim t^{-0.58}$, whereas
$N_c$ is roughly constant with substrate rotation. Both with and without rotation
the dot cross section $A$ finally becomes independent of time, however, it
initially decreases according to $A\sim t^{-0.11}$ with rotation.    

Additionally, for other choices of the sputtering conditions, we find
different patterns which have not been observed experimentally so
far. In particular we found transitions in time from one kind of
surface structures (e.g. smooth, or holes) to other structures (like
non-oriented structures), which can be explained only by the presence
of non-linear effects. Hence, more sputtering experiments with different
ion/substrate types and varying parameters are needed to
verify whether the structures we predict by simulations can indeed be
found experimentally.

\begin{acknowledgments}
This work was funded by the German research association, the Deutsche 
Forchungsgemeinschaft (DFG), within the Sonderforchungsbereich (SFB)
602: {\it Complex Structures in Condensed Matter from Atomic to
  Mesoscopic Scales}. A. K. H acknowledges financial support from the
Volkswagenstiftung within the program ``Nachwuchsgruppen an
Universit\"aten".  
\end{acknowledgments}

\bibliography{pap}

\end{document}